\documentclass[aps,prl,twocolumn]{revtex4-1}

\usepackage{amsmath}
\usepackage{amsfonts}
\usepackage{amssymb}
\usepackage{graphicx}
\usepackage[utf8]{inputenc}
\usepackage[unicode, hidelinks=true]{hyperref}
\usepackage{color}
\usepackage{braket}
\usepackage{dsfont}
\usepackage{MnSymbol}
\usepackage{xspace}
\usepackage{bbold}
\usepackage{diagbox}
\usepackage{setspace}

\setcitestyle{square,numbers}

\DeclareMathOperator{\Tr}{Tr}

\newcommand{\eg}[0]{e.g.\@\xspace}

\newfont{\tensy}{cmsy10}
\newcommand{\chem}[1]{{$\fontdimen16\tensy=3.0pt
    \fontdimen17\tensy=3.0pt \mathrm{#1}$}}

\renewcommand{\S}[0]{\hat{\mathcal{H}}}

\newcommand{\Q}[1]{\hat{Q}_{#1}}
\renewcommand{\P}[1]{\hat{P}_{#1}}

\newcommand{\kF}{k_{\text{F}}}

\newcommand{\kB}{k_{\text{B}}}

\newcommand{\im}{\mathrm{i}}

\newcommand{\absolutetext}[1]{| #1 |}

\newcommand{\expvtext}[1]{\langle #1 \rangle}

\newcommand{\nnpairs}[1]{\langle #1 \rangle}

\newcommand{\spin}[1]{\hat{\mathbf{S}}_{#1}}
\newcommand{\proj}[1]{\hat{\Pi}_{#1}}

\newcommand{\QAFM}{\mathbf{Q}_{\mathrm{AFM}}}
\newcommand{\QVBS}{\mathbf{Q}_{\mathrm{VBS}}}

\newcommand{\psiAFM}{\hat{\mathbf{\Psi}}_{\mathrm{AFM}}}
\newcommand{\psiVBS}{\hat{\Psi}_{\mathrm{VBS}}}

\begin{document}

\title{Valence-Bond Order in a Honeycomb Antiferromagnet Coupled to Quantum Phonons}

\author{Manuel Weber}
\email[Email: ]{mw1162@georgetown.edu}
\affiliation{\mbox{Department of Physics, Georgetown University, Washington, DC 20057, USA}}

\date{\today}

\begin{abstract}
We use exact quantum Monte Carlo simulations to demonstrate that the N\'eel ground
state of an antiferromagnetic SU(2) spin-$\frac{1}{2}$ Heisenberg model on the
honeycomb lattice can be destroyed by a coupling to quantum phonons.
We find a clear first-order transition to a valence-bond-solid state with Kekul\'e order
instead of a deconfined quantum critical point.
However, quantum lattice fluctuations can drive the transition towards weakly first-order,
revealing a tunability of the transition by the retardation of the  interaction.
In contrast to the one-dimensional case,
our phase diagram in the adiabatic regime is qualitatively different from the frustrated $J_1$-$J_2$
model. 
Our results suggest that a coupling to bond phonons can induce Kekul\'e order
in Dirac systems.
\end{abstract}


\maketitle

Exotic phases and phase transitions in quantum many-particle systems have
attracted a lot of interest in the last years.
A recent focus is on valence-bond-solid (VBS) phases in
two-dimensional (2D) spin-$\frac{1}{2}$ antiferromagnets (AFMs)
where translational symmetry is spontaneously broken via the formation of dimers
between neighboring spins \cite{Sachdev:2008aa}.
The proliferation of topological defects in the AFM/VBS order parameter
\cite{PhysRevLett.62.1694,PhysRevB.42.4568,PhysRevB.70.220403}
has been proposed to drive a continuous quantum phase transition between the two phases.
The scenario of a deconfined quantum critical point (DQCP) \cite{Senthil1490,PhysRevB.70.144407}
is beyond the Landau-Ginzburg-Wilson paradigm
where competing orders with different broken symmetries 
require a first-order transition.
Furthermore, the interplay between the topological defects
of the VBS phase and disorder is 
currently explored
\cite{PhysRevX.8.031028}.

VBS order often appears in frustrated spin models,
but their numerical study in 2D
is usually restricted to small
system sizes
or approximate schemes.
Large-scale
quantum Monte Carlo (QMC) simulations give exact results for a class of sign-problem free
Hamiltonians called $J$-$Q$ models \cite{PhysRevLett.98.227202}
that are specifically designed to generate the desired orders.
While $J$-$Q$ models show strong evidence for a continuous AFM--VBS transition---%
most notably on the square lattice
\cite{PhysRevLett.98.227202,
PhysRevLett.100.017203,
PhysRevLett.104.177201}%
---%
the scenario of a weak first-order transition cannot be completely ruled out
\cite{Jiang_2008,PhysRevLett.101.050405}.
Recently, unconventional first-order transitions with enhanced symmetry have been reported \cite{Zhao:2019aa}.
It is of current interest to find VBS phases also in more realistic models
beyond designer Hamiltonians.

In quasi-1D systems such as the organic TTF compounds \cite{PhysRevLett.35.744}
or the inorganic material
\chem{CuGeO_3} \cite{PhysRevLett.70.3651}, VBS order often arises from
the spin-Peierls instability \cite{PhysRevB.10.4637, PhysRevB.19.402},
which is closely related to a $2\kF$ Fermi-surface instability in
 electronic models.
A 1D Heisenberg model is unstable towards dimerization
for any finite coupling to classical phonons,
because the gain in magnetic energy is  higher
than the loss in elastic energy.
However, quantum lattice fluctuations can stabilize a gapless phase with critical AFM correlations below a critical coupling.
The phase diagrams of 1D spin-phonon models have been determined
numerically 
\cite{Augier:1998aa, PhysRevB.58.9110, PhysRevLett.81.3956,
PhysRevLett.83.408, PhysRevB.74.214426,
PhysRevB.56.14510, PhysRevB.60.12125, PhysRevLett.83.195, 2007arXiv0705.0799M, PhysRevLett.115.080601}.
For high phonon frequencies,
the spin-Peierls problem
maps to
the frustrated $J_1$-$J_2$ model
with next-nearest neighbor Heisenberg exchange
\cite{doi:10.1143/JPSJ.56.3126,PhysRevB.57.R14004, PhysRevB.60.6566, PhysRevB.65.144438}.
In particular, the quantum phase transition at finite phonon frequencies is in the same
universality class as in the $J_1$-$J_2$ model \cite{PhysRevLett.115.080601}.

The relevance of spin-phonon interactions 
in 1D
is acknowledged by the fact that---even in other contexts---%
the VBS state is sometimes called \textit{spin-Peierls state}
\cite{sachdev_2011,PhysRevLett.62.1694,PhysRevB.42.4568}.
By contrast, the nature and even the existence of VBS order
in 2D is still under debate and has only been explored on the square lattice.
The spin-Peierls model was initially studied in the context
of high-$T_\mathrm{c}$ superconductivity as the large-$U$ limit of the
Peierls-Hubbard model.
Different
dimerization patterns were discussed
as the ground-state configurations of classical phonons
\cite{PhysRevB.37.1569, PhysRevB.37.9546,PhysRevB.39.12324,PhysRevB.65.085102,
PhysRevB.65.134409,PhysRevB.74.014425,doi:10.1143/JPSJ.75.034705},
even a resonating valence-bond state was proposed
\cite{ANDERSON1196,PhysRevB.35.8865}. The stability
of the spin-Peierls state
was questioned because a large Hubbard repulsion favors AFM order and suppresses VBS order
in 2D \cite{PhysRevB.36.7190}.
So far, exact numerical simulations were inhibited
by the large bosonic Hilbert space and difficult phonon sampling.
The only available QMC study which approached the full 
quantum-phonon problem
did not find VBS order 
\cite{PhysRevB.68.184416}.

\begin{figure}
  \includegraphics[width=\linewidth]{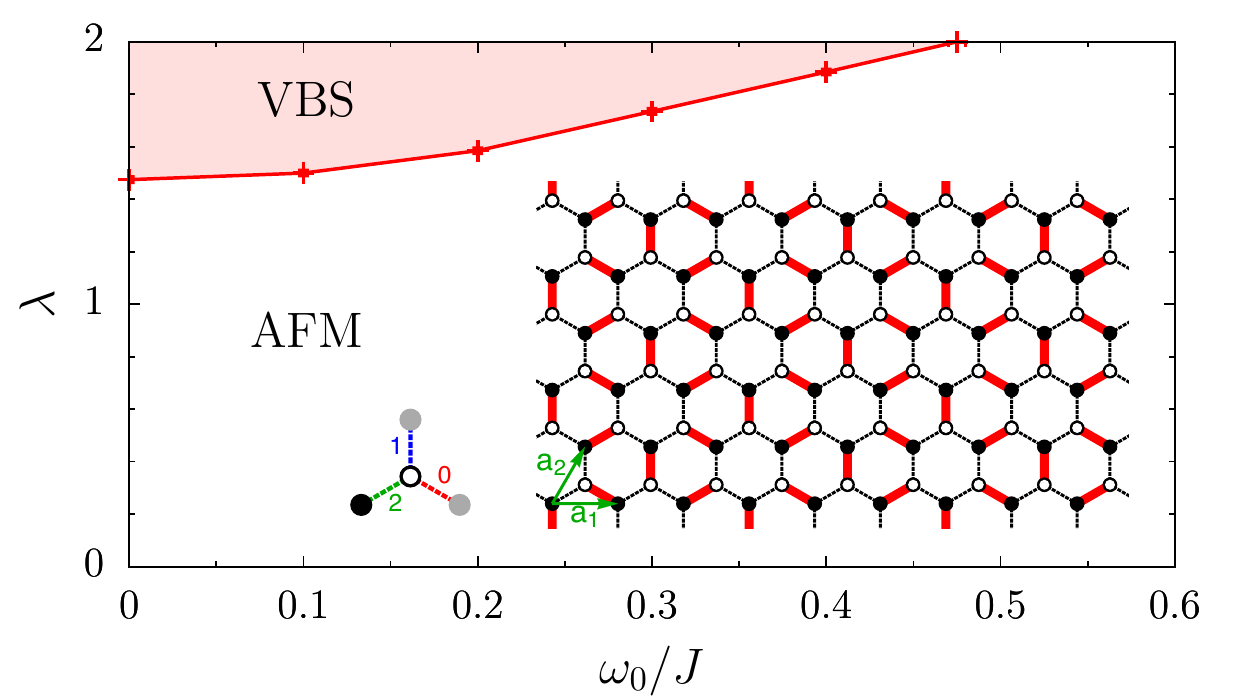}
  \caption{\label{fig:phase_diagram}
Phase diagram of the spin-Peierls model~(\ref{Eq:Hamiltonian})
as a function
of phonon frequency and spin-phonon coupling from QMC simulations.
The large inset shows a honeycomb lattice with columnar VBS order
where strong (weak) links represent a high (low) $\expvtext{\proj{ij}}$.
The small inset depicts the two sites $A,B$
and the three bonds $\mu=0,1,2$ that belong to a unit cell.
  }
\end{figure}

The honeycomb lattice has a lower coordination number than the square lattice
which makes the VBS state energetically more favorable.
A columnar VBS state with Kekul\'e order
(see inset of Fig.~\ref{fig:phase_diagram}) was found in a
$J$-$Q$ model and its AFM--VBS transition was interpreted in
terms of a DQCP \cite{PhysRevLett.111.087203, PhysRevLett.111.137202, PhysRevB.88.220408, PhysRevB.91.104411}.
Similar  transitions appear in
Dirac systems \cite{PhysRevLett.119.197203,2019arXiv190410975L} 
where the emergence of Kekul\'e order is a recent focus of theoretical
\cite{PhysRevLett.98.186809, PhysRevB.80.205319,
PhysRevB.82.035429,
Li:2017aa, PhysRevB.94.205136, PhysRevB.96.115132,
PhysRevB.98.121406,
PhysRevLett.123.157601}
and experimental \cite{Gomes:2012aa, 2016NatPh..12..950G} studies.
Interaction effects in graphene 
have attracted 
additional interest
since the discovery of superconductivity in twisted bilayer graphene \cite{Cao:2018aa}.

In this Letter, we demonstrate
that spin-phonon coupling can stabilize a columnar VBS state and determine the
ground-state
phase diagram of the spin-Peierls model as a function of phonon
frequency
(see Fig.~\ref{fig:phase_diagram}).
Our simulations were made possible by a recently developed QMC method
that solves the full quantum-phonon problem efficiently using retarded interactions
\cite{PhysRevLett.119.097401}.
The AFM--VBS transition is strongly first-order for classical phonons,
but quantum lattice fluctuations can drive the transition towards weakly first-order.
We discuss 
how our results are
related to the putative DQCP scenario on the honeycomb lattice.
Furthermore, we debate whether retardation effects can induce the
physics of the frustrated $J_1$-$J_2$ model at high phonon frequencies which are
not accessible to our simulations. At low frequencies, the two models show different orders,
unlike in the 1D case.
Finally, our results suggest that a coupling to bond phonons 
can induce Kekul\'e order in Dirac systems.

\textit{Model \& Method.}---We consider the spin-Peierls model
\begin{align}
\label{Eq:Hamiltonian}
\hat{H}
	=
	\sum_{\nnpairs{i,j}} \left(J + \alpha \, \Q{ij} \right) \spin{i} \cdot \spin{j}
	+ \sum_{\nnpairs{i,j}} \left( \mbox{$\frac{1}{2M}$} \P{ij}^2 +  \mbox{$\frac{K}{2}$} \Q{ij}^2 \right)
\end{align}
where the antiferromagnetic exchange $J$ is modulated via a coupling $\alpha$
to optical bond phonons with frequency $\omega_0 = \sqrt{K/M}$.
The spin-$\frac{1}{2}$ operators $\spin{i}$ are defined on the sites $i$ of a honeycomb lattice, whereas  the phonon momenta
$\P{ij}$ and displacements $\Q{ij}$ act on the links between nearest neighbors $\nnpairs{i,j}$.
In the following, we use $J=1$ as the unit of energy, define
the dimensionless coupling $\lambda=\alpha^2/(2KJ)$, and set $\hbar, \kB =1$.

The phonons can be integrated out exactly using the imaginary-time path integral.
The partition function becomes
$Z =
Z_0
\Tr \hat{\mathcal{T}}_\tau \, e^{-\S}$
with $\S = \S_{J} +\S_{\lambda}$ and
\begin{align}
\label{eq:SJ}
\S_{J}
	&=
	 - J'  \int_0^\beta d\tau \sum_{\nnpairs{i,j}} \proj{ij}(\tau) \, ,
	 \qquad
	 J' = J\left(1 - \frac{\lambda}{2} \right) \, , \\
\S_\lambda
	&=
	- \lambda J \iint_0^\beta d\tau d\tau' \sum_{\nnpairs{i,j}}
	\proj{ij}(\tau) \, P(\tau-\tau') \, \proj{ij}(\tau') \, .
\label{eq:Slam}
\end{align}
The spin-phonon coupling leads to a retarded interaction $\S_\lambda$ between
singlet projectors $\proj{ij} = \frac{1}{4} -  \spin{i} \cdot \spin{j}$
at different times $\tau$, $\tau'$
and is mediated by the free-phonon propagator
$P(\tau)= e^{-\omega_0 \tau} \omega_0 / (1-e^{-\omega_0 \beta})$.
Because $ \spin{i} \cdot \spin{j}$ is shifted by $\frac{1}{4}$,
the Heisenberg exchange $J'$ gets renormalized with $\lambda$.
Here, $\beta = 1/T$ is the inverse temperature and $Z_0$ includes 
the partition function of free phonons.

For our simulations we used a recently developed QMC method for
retarded interactions \cite{PhysRevLett.119.097401} that is based on
a diagrammatic expansion of $Z/Z_0$ in $\S$.
The method is closely related to the 
stochastic series expansion
\cite{PhysRevB.43.5950}
and makes use of efficient directed-loop updates \cite{PhysRevE.66.046701}.
It only has statistical errors and is
free of a sign problem for $\lambda \leq 2$ ($J' \geq 0$).
The use of retarded interactions avoids the difficulties of direct phonon
sampling which inhibited previous studies of the 2D case,
but system sizes are still limited by the generically difficult sampling near a 
first-order transition.
We use an exchange Monte Carlo method \cite{1996JPSJ...65.1604H, PhysRevB.65.155113}
to improve simulations in the VBS phase.
Phonon observables can be recovered from the perturbation expansion
using generating functionals \cite{PhysRevB.94.245138}.
Further details on our method are presented elsewhere~\cite{SpinPeierlsMethod}.

Simulations were performed on $L\times L$ honeycomb lattices with $2L^2$ spins
and periodic boundary conditions.
We used $\beta J = 2L$ which is suitable for detecting
ground-state order of a continuous phase transition with dynamical
exponent $z=1$ or a first-order transition.

\textit{Results.}---%
The phase diagram in Fig.~\ref{fig:phase_diagram} contains AFM and VBS phases
which can be identified from a finite-size analysis
of the (basis-dependent)
order parameters
\cite{PhysRevB.91.104411}
\begin{align}
\label{Eq:orderp_AFM}
\psiAFM(\mathbf{q})
&=
\frac{1}{2L^2} \sum_{\mathbf{r}}
\left( \spin{\mathbf{r}A} - \spin{\mathbf{r}B} \right)
e^{\im \mathbf{q} \cdot \mathbf{r}} \, , \\
\psiVBS(\mathbf{q})
&=
\frac{1}{2L^2} \sum_{\mathbf{r}}
\sum_{\mu=0}^2 \proj{\mathbf{r}\mu} \, e^{2\pi \im \mu /3} \,
e^{\im \mathbf{q} \cdot \mathbf{r}} \, .
\label{Eq:orderp_VBS}
\end{align}
Here, $\mathbf{r}$ is the position vector of the Bravais lattice. Each unit cell
has two sites $A,B$ and three bonds $\mu=0,1,2$ which are chosen
as depicted in Fig.~\ref{fig:phase_diagram}.
AFM order breaks 
the SU(2) spin
symmetry and appears at
 $\QAFM = (0,0)$, whereas the columnar VBS state breaks a $Z_3$ lattice symmetry
such that spin singlets are arranged in a Kekul\'e pattern with $\QVBS = (2\pi/3,-2\pi/3)$,
as shown in Fig.~\ref{fig:phase_diagram}.
We measure 
$C_{\alpha}(\mathbf{q}) = \expvtext{\absolutetext{\hat{\Psi}^z_\alpha(\mathbf{q})}^2}$
after replacing $\spin{i} \to \hat{S}_i^z$ in Eqs.~(\ref{Eq:orderp_AFM}) and (\ref{Eq:orderp_VBS})
to calculate the correlation ratios \cite{Binder:1981aa}
\begin{align}
R_\alpha
	=
	1 - \frac{C_\alpha(\mathbf{Q}_\alpha + \delta \mathbf{q})}{C_\alpha(\mathbf{Q}_\alpha)}
\end{align}
with $\delta\mathbf{q}=(0,2\pi/L)$.
When $L\to \infty$, $R_\alpha(L) \to 1$ in the corresponding
ordered phase and $R_\alpha(L) \to 0$ in the disordered phase.
The same holds for the Binder cumulant
$
U_\mathrm{VBS}
= 2 -
\expvtext{\absolutetext{\hat{\Psi}^z_\mathrm{VBS}(\mathbf{Q}_\mathrm{VBS})}^4}
/
\expvtext{\absolutetext{\hat{\Psi}^z_\mathrm{VBS}(\mathbf{Q}_\mathrm{VBS})}^2}^2
$.

\begin{figure}
  \includegraphics[width=\linewidth]{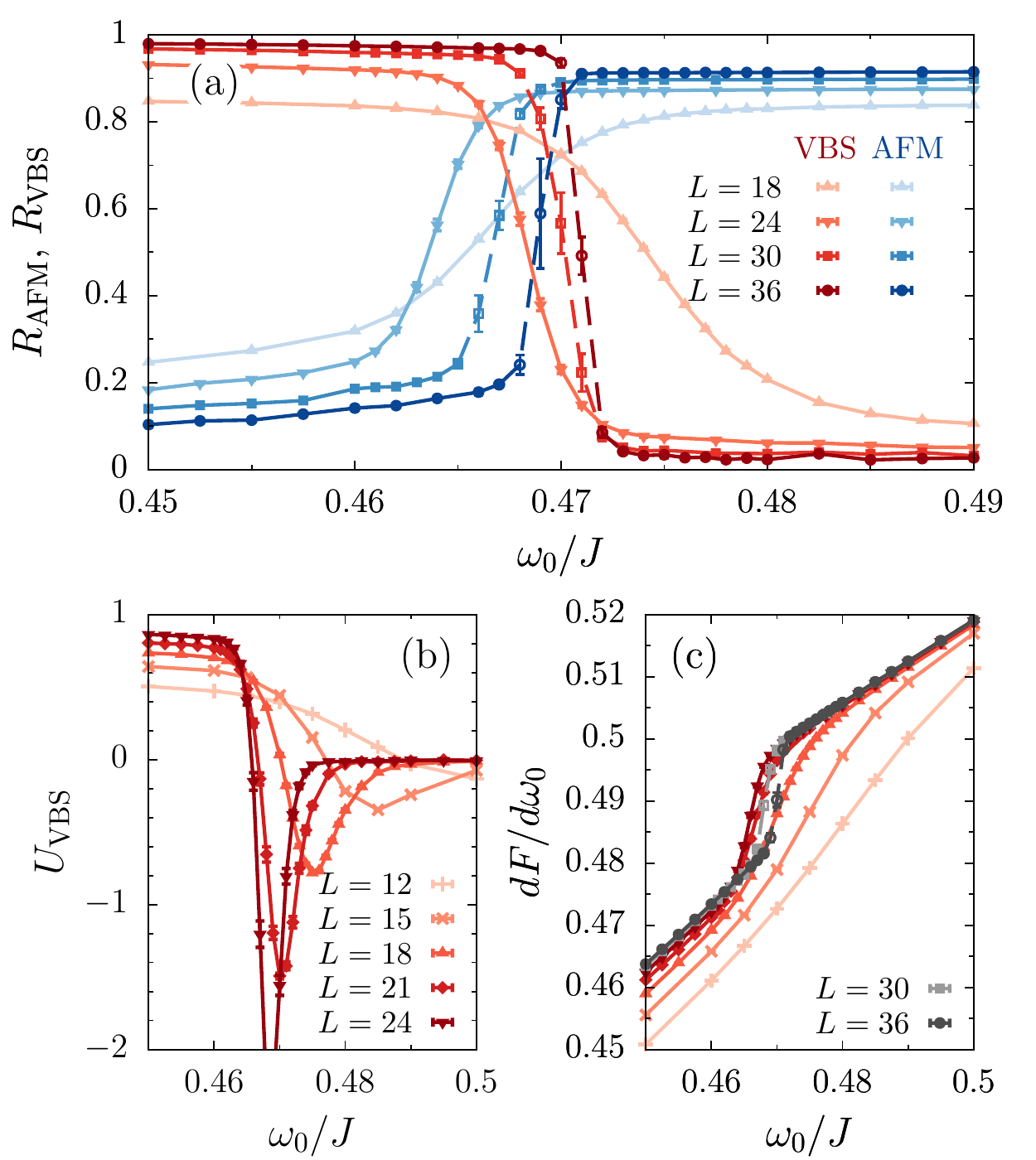}
  \caption{\label{fig:correlation_ratios}
Finite-size analysis of the AFM--VBS transition at $\lambda=2$ as a function of $\omega_0$.
(a) AFM/VBS correlation ratios, (b) VBS Binder ratio, and (c) free-energy derivative
$dF/d\omega_0$.
Labels in (b) also apply to (c).
Open symbols and dashed lines represent data points
where the tunneling times between coexisting orders are longer than our simulation times.
  }
\end{figure}

Figure \ref{fig:correlation_ratios} shows our results for the retardation-driven AFM--VBS transition
at $\lambda=2$.  Both orders can be identified from the correlation ratios
in Fig.~\ref{fig:correlation_ratios}(a)
which indicate
a sharp transition at
$\omega_{0,\mathrm{c}} /J \approx0.47$.
The Binder ratio in Fig.~\ref{fig:correlation_ratios}(b) develops a negative peak
that diverges with $L$---%
a typical finite-size effect at a first-order transition and a result of
phase coexistence separated by an energy barrier 
\cite{Vollmayr:1993aa}. 
Further evidence is given by an emerging discontinuity in the free-energy derivative $dF/d\omega_0$
in Fig.~\ref{fig:correlation_ratios}(c).
A precise extrapolation of $\omega_{0,\mathrm{c}}$ is complicated by the nonmonotonic
drift of finite-size estimates towards lower (higher) $\omega_0$ for $L<24$ ($L>24$)
as well as difficult Monte Carlo sampling 
in the coexistence region.

\begin{figure}[b]
  \includegraphics[width=0.95\linewidth]{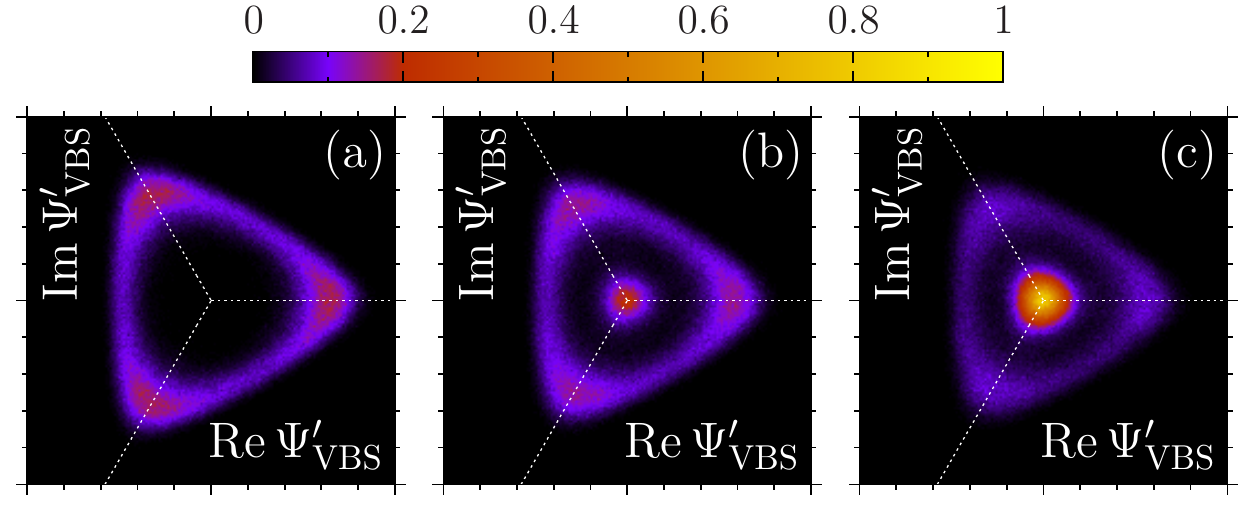}
  \caption{\label{fig:histo}
Histograms of the VBS order parameter ${\hat{\Psi}'_\mathrm{VBS}}$ across
the AFM--VBS transition for
(a) $\omega_0/J =0.45$, (b) $\omega_0/J = 0.466$, and
(c) $\omega_0/J=0.47$.
Here, $L=18$ and $\lambda=2$.
  }
\end{figure}

The nature of the VBS phase is not entirely determined by its ordering vector.
Besides the columnar VBS state illustrated in Fig.~\ref{fig:phase_diagram},
$\QVBS=(2\pi/3,-2\pi/3)$  can also correspond to a plaquette VBS state where
strong and weak links are interchanged. The two states are
 distinguished by the phase of the complex VBS order parameters \cite{PhysRevB.91.104411}.
We consider the modified field $\hat{\Psi}'_\mathrm{VBS}$ by replacing 
$\proj{\mathbf{r}\mu} \to \left(J + \alpha \, \Q{\mathbf{r}\mu} \right)  \proj{\mathbf{r}\mu}$ in Eq.~(\ref{Eq:orderp_VBS}) because
its expectation value
$
{\Psi}'_\mathrm{VBS}
=
{(2L^2 \beta)^{-1}}
	\sum_{\mathbf{r}\mu}
	\expvtext{n({\proj{\mathbf{r\mu}}})}_\mathrm{MC}\,
	e^{2\pi\im \mu/3} \, e^{\im \QVBS \cdot \mathbf{r}}
$
can be easily estimated from the number of $\proj{\mathbf{r\mu}}$
per Monte Carlo configuration \cite{PhysRevLett.98.227202,PhysRevB.94.245138}.
The histogram of $\Psi'_\mathrm{VBS}$ in Fig.~\ref{fig:histo}(a)
illustrates that VBS order appears at the columnar angles $e^{2\pi\im \mu/3}$.
The emergence of a central peak in Figs.~\ref{fig:histo}(b) and \ref{fig:histo}(c)
indicates coexisting AFM order.
Moreover, the threefold anisotropy of the VBS order parameter remains robust
in the coexistence region.

\begin{figure}
  \includegraphics[width=\linewidth]{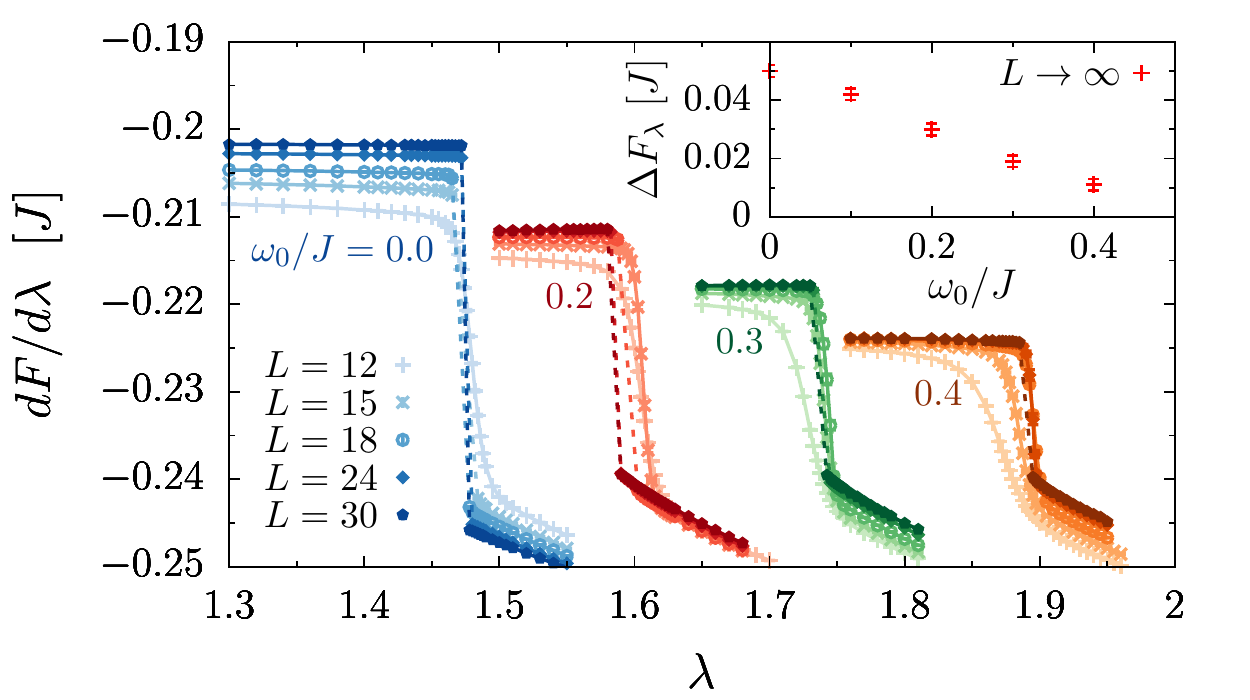}
  \caption{\label{fig:dFdlam}
Free-energy derivative as a function of $\lambda$ for different $\omega_0$ and $L$.
The inset shows the jump at the critical coupling extrapolated to $L\to \infty$.
Color scheme based on Ref.~\cite{2014zndo.....10282S}.
  }
\end{figure}

Figure~\ref{fig:dFdlam} shows the free-energy derivative $dF/d\lambda$ for
different $\omega_0$.
The critical couplings
in Fig.~\ref{fig:phase_diagram}
are determined from the discontinuities in $dF/d\lambda$ and increase with increasing $\omega_0$.
The strength of a first-order transition can be 
characterized by the size of the jump in its free-energy derivative.
To estimate $\Delta F_\lambda(L)$, we extrapolate
the two branches of $dF/d\lambda$ towards the center of the coexistence region.
A final extrapolation $L\to \infty$ leads to the jumps summarized in the inset of  Fig.~\ref{fig:dFdlam}.
We find that the transition is 
significantly weakened
with increasing $\omega_0$.

\textit{Discussion.}---%
The nature of the AFM to columnar VBS transition on the honeycomb lattice
has been studied numerically in the $J$-$Q$ model
\cite{PhysRevLett.111.087203, PhysRevLett.111.137202, PhysRevB.88.220408, PhysRevB.91.104411}. A finite-size analysis obtained critical exponents with logarithmic violations
of scaling, consistent with the interpretation of a continuous transition on the square lattice \cite{PhysRevLett.104.177201}.
However, instead of showing an emergent U(1) symmetry at criticality \cite{Senthil1490}, 
as observed on the square lattice \cite{PhysRevLett.98.227202},
$\psiVBS(\QVBS)$ retained a 3-fold anisotropy 
which was interpreted in terms of near-marginal behavior of the topological
defects \cite{PhysRevLett.111.087203}. Conformal bootstrap as well as
an analysis of anomalies
in corresponding field theories
suggested that 3-fold monopoles
are slightly relevant at criticality in SU(2) spin models
on the honeycomb lattice
\cite{PhysRevB.92.184413, PhysRevLett.117.131601, PhysRevB.98.085140},
but that lattice sizes of $L \leq 72$ \cite{PhysRevLett.111.087203}
or $L \leq 96$ \cite{PhysRevB.88.220408}
were too small to find evidence for a weak first-order transition
in the $J$-$Q$ model.
The spin-Peierls model studied in this Letter serves as an example where
the AFM--VBS transition is clearly first-order already on small system sizes
and therefore follows the Landau-Ginzburg-Wilson paradigm.
The different length scales seem to originate from the retarded nature of
$\S_\lambda$.
The first-order
transition 
is strongest at $\omega_0 = 0$
where the nontrivial minimization of $\expvtext{\hat{H}(Q_{ij})}$
in terms of the real-valued static displacements
only permits certain ordering patterns \cite{PhysRevLett.107.066801}.
Here, the interaction range in time,
$P(\tau) \sim e^{-\omega_0 \tau}$,
is largest, but the transition is significantly weakened with increasing $\omega_0$.
In $J$-$Q$ models, singlet projectors $\proj{ij}$ interact
at equal times but between different bonds of the lattice to induce VBS order.

Our numerical study is restricted to $\lambda \leq 2$
due to a sign problem, but it is worth speculating on how the phase
diagram in Fig.~\ref{fig:phase_diagram} continues for $\omega_0/J > 0.5$.
We expect that $\omega_0$ tunes the AFM--VBS transition towards weakly first-order,
as the discontinuity of $dF/d\lambda$ in Fig.~\ref{fig:dFdlam} tends to further decrease.
A reliable extrapolation of $\Delta F_\lambda(\omega_0)$
is out of reach, but it also
seems possible that $\Delta F_\lambda$ 
vanishes or that new physics arises
at higher $\omega_0$.
In the limit $\omega_0\to\infty$, the spin-Peierls model maps to a Heisenberg model
with AFM order only, 
but
for small 
interaction ranges
$1/\omega_0$ in time,
the retardation in
$\S_\lambda$
effectively generates
longer-range 
spin interactions,
but also
higher-order 
corrections
\cite{doi:10.1143/JPSJ.56.3126,PhysRevB.57.R14004}.
Such a mapping to the frustrated $J_1$-$J_2$ model successfully describes the physics of the 1D spin-Peierls model
\cite{PhysRevB.60.6566, PhysRevB.65.144438}.
The complex phase diagrams 
of frustrated spin chains 
can also be found
in electron-phonon models
where $\omega_0$
drives the 
competition between different $2\kF$ orders
separated by a 1D DQCP \cite{PhysRevResearch.2.023013}.
On the honeycomb lattice,
the $J_1$-$J_2$ model
has been studied on small clusters
using exact diagonalization \cite{Mosadeq_2011, PhysRevB.84.024406}
and the 
density-matrix renormalization group
\cite{PhysRevLett.110.127203,PhysRevLett.110.127205,PhysRevB.88.165138}.
As a function of increasing ratio $J_2/J_1$, these studies found AFM, plaquette VBS, and staggered
VBS order.
The AFM--VBS transition was interpreted in terms of a DQCP 
\cite{PhysRevB.84.024406,PhysRevLett.110.127203,PhysRevLett.110.127205,PhysRevB.88.165138},
whereas an intermediate spin-liquid phase had also been discussed \cite{PhysRevB.96.104401}.
Whether the physics of the $J_1$-$J_2$ model appears at high $\omega_0$,
depends on two questions:
(i)~How do the effective nearest- and next-nearest-neighbor 
couplings 
$J_{1,2}(\omega_0,\lambda)$
depend on the parameters
of the the spin-Peierls model? 
In particular, will they reach a nontrivial  regime 
in the phase diagram beyond AFM order?
(ii)~Do other operators become relevant in the mapping?
The latter must be true
for $\omega_0/J < 0.5$.
Although VBS order
appears in both models at 
$\QVBS=(2\pi/3,-2\pi/3)$,
our results show columnar instead of plaquette order.
Therefore, the adiabatic regime $\omega_0 \ll J$ is not described
by the $J_1$-$J_2$ model.
While this is not surprising because the mapping should only  hold
at high $\omega_0$, 
the 1D
problem is
governed by 
the $J_1$-$J_2$ model
even at frequencies as low as
$\omega_0/J = 0.25$ \cite{PhysRevLett.115.080601}.
As there is only one possible VBS pattern in 1D,
 the nature of the VBS phase cannot change with $\omega_0$.
Whether quantum lattice fluctuations can change the ground-state physics in 2D, remains open.

Our results on the honeycomb lattice demonstrate that spin-phonon coupling can induce
VBS order in a 2D antiferromagnet.
Although a previous QMC study did not find a
VBS phase on the square lattice \cite{PhysRevB.68.184416},
it is likely to exist in the regime $\lambda>2$ not
accessible to simulations.
While spin-phonon interactions are a relevant mechanism
in materials, the critical couplings found in this Letter are rather strong,
as it is also the case in many other spin models, \eg, the $J$-$Q$ models \cite{PhysRevLett.98.227202}.
A coupling to phonons was found to be important in combination
with frustration \cite{PhysRevLett.89.037204}, \eg, on
hexagonal \cite{PhysRevLett.103.067204}
or pyrochlore lattices \cite{PhysRevLett.88.067203,PhysRevLett.93.197203}.
Moreover, 
the spin-Peierls model is closely related to electron-phonon models:
 it corresponds to a Su-Schrieffer-Heeger (SSH) model \cite{PhysRevLett.42.1698} with
infinite Hubbard repulsion. 
Only recently, 
determinantal QMC studies of the 2D SSH model with
quantum phonons were carried out \cite{Li:2020aa, 2020arXiv200509673X},
but available system sizes 
were restricted by the difficult phonon sampling.
On the square lattice,
the SSH model supports VBS order at $\QVBS=(\pi,\pi)$
\cite{2020arXiv200509673X},
whereas the influence of the Hubbard repulsion is still under debate
\cite{PhysRevB.37.9546, PhysRevB.65.134409}.
On the honeycomb lattice, QMC results are only available for Holstein
phonons which lead to a charge-density-wave phase
\cite{PhysRevLett.122.077601,PhysRevLett.122.077602}.
Kekul\'e order was proposed to appear from a coupling
to SSH phonons \cite{PhysRevLett.103.216801, PhysRevB.85.155439, PhysRevB.90.035122}.
Recently, the SSH-Hubbard model was studied in the limit $\omega_0 \to \infty$,
where a direct (DQCP) transition between columnar VBS
and AFM order was reported~\cite{2019arXiv190410975L}, as well as
a fermion-induced quantum critical point
between a Dirac semimetal and VBS order \cite{Li:2017aa}.
Our results in the large-$U$ limit suggest that Kekul\'e order
and the corresponding transitions also exist at finite $\omega_0$
but the AFM--VBS transition might turn first-order for low $\omega_0$.

\textit{Conclusions \& Outlook.}---%
We demonstrated
that VBS order can arise in a spin-$\frac{1}{2}$ Heisenberg model
 coupled to phonons.
The first-order transition from AFM to columnar VBS order
disagrees with the putative DQCP scenario on the honeycomb lattice,
but can be tuned towards weakly first-order when
quantum lattice fluctuations become stronger.
Our results establish retardation effects as an important influence
on the AFM--VBS transition
that was
not considered
in previous studies.
Our recently developed 
QMC method for retarded interactions \cite{PhysRevLett.119.097401}
enables future work in this direction. 
In particular, it seems possible
to engineer different orders via an appropriate
couplings to phonons and thereby extend the zoo of models
that show nontrivial phases in sign-problem-free QMC simulations.
While retardation is an established mechanism to induce frustrated
interactions in 1D models, the columnar VBS order at $\omega_0/J <0.5$
is in contrast to 
the plaquette VBS order found in the $J_1$-$J_2$ model
\cite{Mosadeq_2011, PhysRevB.84.024406,PhysRevLett.110.127203,PhysRevLett.110.127205,PhysRevB.88.165138}.
It remains an open question whether the 2D spin-Peierls model displays
the phases of the $J_1$-$J_2$ model or any
other nontrivial physics at higher $\omega_0$.
Moreover, it will be of interest to explore
how thermally-generated phonon fluctuations
modulate the exchange integral $J_{ij}(\Q{ij})$
in the VBS phase and lead to a disordered phase.
Finally, the possibility of finding Kekul\'e order in
Dirac or spin
systems motivates future studies of phonon coupling.

\begin{acknowledgments}
  \textit{Acknowledgments.}---%
  We thank  F.~Assaad, J.~Freericks,
  M.~Hohenadler, and F.~Parisen Toldin for helpful discussions.
  This work was supported by the U.S. Department of Energy (DOE),
  Office of Science, Basic Energy Sciences (BES) under Award
  DE-FG02-08ER46542. 
The authors gratefully acknowledge the Gauss Centre for Supercomputing e.V. (www.gauss-centre.eu) for funding this project by providing computing time on the GCS Supercomputer SuperMUC-NG at Leibniz Supercomputing Centre (www.lrz.de) (project-id pr53ju).
\end{acknowledgments}

%


\end{document}